# INAF-Osservatorio astrofisico di Torino
*Technical Report nr. 178*

# The slowly variable star FY Lacertae

*A. Fucci, F. Maldarella, D. Pace, V. A. Zasa Courtial,* **C. Benna, D. Gardiol**, **G. Pettiti**

*Pino Torinese, 19 ottobre 2017*

# THE SLOWLY VARIABLE STAR FY LACERTAE

FUCCI ALICE[1]; MALDARELLA FABIO[1]; PACE DAVIDE[1] AND ZASA COURTIAL VITTORIO AUGUSTO[1]
BENNA CARLO[2]; GARDIOL DANIELE[2] AND PETTITI GIUSEPPE[2]

1) Curie StemLab group, IIS Curie Vittorini, Corso Allamano 130, 10095, Grugliasco(TO), Italy, mylabcurie@gmail.com
2) Istituto Nazionale di Astrofisica, Osservatorio Astrofisico di Torino, via Osservatorio 20, I-10025 Pino Torinese (TO) – Italy, pettitg@alice.it

**Abstract:** Photometric observations and analysis of FY Lac were performed in order to determine the variability characteristics, using measurements taken at Loiano site of Bologna Astronomical Observatory and AAVSO data. The star shows a Long Term variability with a preliminary period of about 274 ± 28 days and photometric characteristics compatible with a red M5 III giant star, with a temperature of 3420 °K.

## 1  Introduction

FY Lac is a pulsating-variable star currently included in the list of variable stars under investigation at Astronomical Observatory of Torino (OATo). It was previously studied by Hoffmeister (1966) who defined it as a red slowly variable star.
This report provides results of recent photometric observations in B-V Johnson-Cousins standard filters, carried on at Loiano site Astronomical Observatory of Bologna. It also analyzes not transformed data available from AAVSO database.
FY Lac is also known as BD+44 4159, IRAS 22304+4516, 2MASS J22323551+4531461 or TYC 3620-449-1 and is located at R.A. = 22h 32m 35.50s , DEC = +45° 31' 46.3" (ICRS J2000).

## 2  Observation

The photometric observations were carried out at Loiano site of the Bologna Astronomical Observatory (OABo).
The instrumentation characteristics are reported in Tab. 1.

| Loiano site of the Bologna Astronomical Observatory (OABo) | | | | | | |
|---|---|---|---|---|---|---|
| **Telescope (Cassini)** | | | **Detector (BFOSC)** | | | |
| Useful Diameter | Focal Length | Optical conf. | Camera | Array | Johnson-Kron-Cousins | FoV |
| 150.0 cm | 1200 cm | Ritchey-Chrétien | EEV D129915 | 1300x1340 pixels | B, V | 13'x12.6' |

**Tab. 1- Loiano site instrumentation characteristics**

The number of observations and length of exposures in each filter is shown in Tab. 2.

| | | Number of observations - Length of exposure (s) | |
|---|---|---|---|
| **Date** | **Time span (min.)** | **B** | **V** |
| 7 Dec, 2016 | 35 min. | 20 - 20s | 20 - 6s |

**Tab. 2- Observation log**

The data of the comparison star and check star are defined by AAVSO finding chart X16317OS and are shown in Tab. 3. A finding chart with the identification of the stars is in Fig. 1.

| Star | ID | RA [h m s] (J2000.0) | DEC [° ´ ´´] (J2000.0) | B | V |
|---|---|---|---|---|---|
| Comparison | TYC 3620-991-1 | 22 32 58.28 | +45 32 15.64 | 11.678 ± 0.064 | 10.985 ± 0.025 |
| Check | TYC 3620-919-1 | 22 33 13.01 | +45 33 13.30 | 12.489 ± 0.071 | 11.669 ± 0.032 |

**Tab. 3- Comparison and check star data**

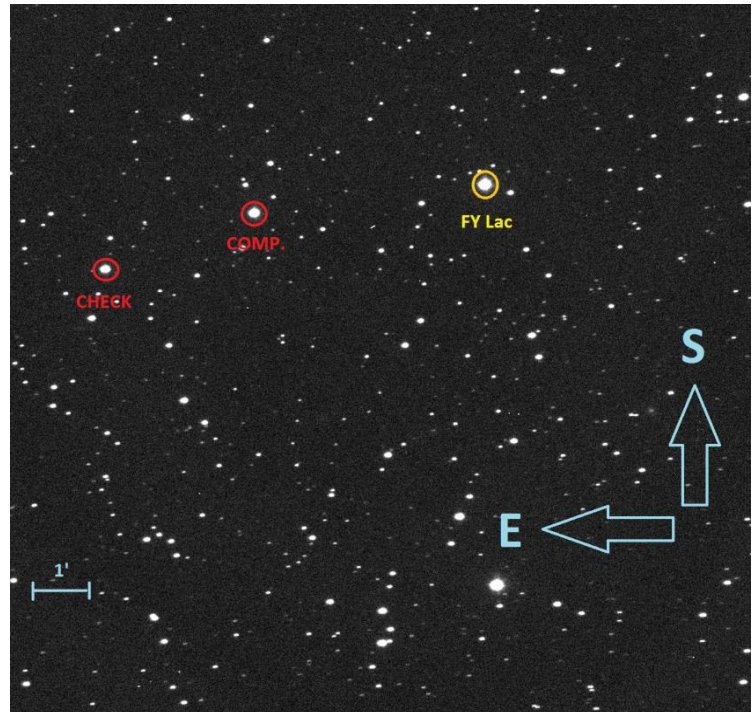

**Fig. 1- FY Lac, the comparison and check stars.**

For all 40 images, Dark and Flat Field corrections were applied and aperture differential photometry was performed using the software AstroArt 5.0.

The basic equation applied to transform the instrumental magnitudes to standard magnitudes is defined, for filter V, in Eq. 1:

$$Vvar = \Delta v + Tv * \Delta(B-V) + Vcomp \qquad (1)$$

where:
- $\Delta v$ is the instrumental magnitude of the variable minus the instrumental magnitude of the comparison star (vvar- vcomp);
- Vcomp is the V–magnitude of the comparison star defined in Tab. 4;
- Tv is the transformation coefficient defined in Tab. 5;

- Δ(B-V) is the difference in the standard color of the variable versus the standard color of the comparison star, computed using the formula:

$$\Delta(B\text{-}V) = T_{bv} * \Delta(b\text{-}v)$$

Similar equation was applied for the filter B.

In addition, we analyzed not transformed measurements available from the AAVSO database.

## 3   Data analysis

Not transformed data available from AAVSO database were analyzed to identify medium-long term variability of the star using version 2.51 of the light curve and period analysis software PERANSO.

The accuracy of the instrumental magnitudes was calculated based on C-K deviations standard and is: 0.005 in B and 0.004 in V filter.

The accuracy of the standard magnitudes was calculated using a propagation error approach based on comparison star accuracy, defined in Tab. 3.

## 4   Results

The B, V standard magnitude and B-V colour index calculated from the data of the Loiano Astronomical Observatory are shown in Figure 2, 3 and 4.

The tables represent Standard Magnitudes vs. HJD middle time exposure.
We found the following mean values for B, V and B-V:

- B = 11.897 ± 0.066;
- V = 10.097 ± 0.025;
- B-V = 1.800 ± 0.070.

**Fig. 2- Standard Magnitude filter B**

**Fig. 3- Standard Magnitude filter V**

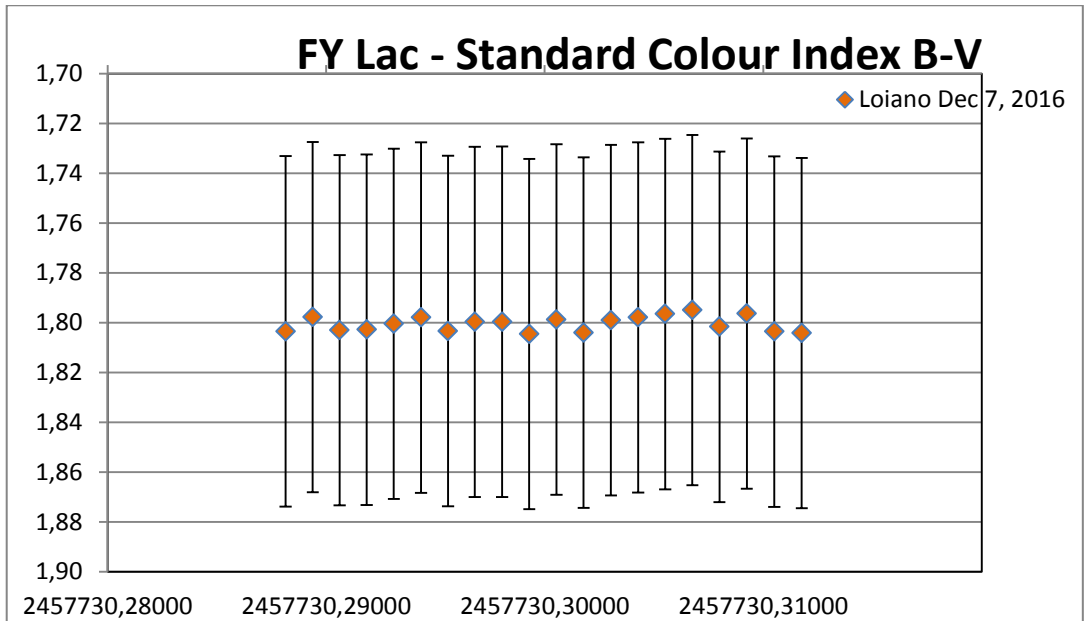

**Fig. 4- Standard Colour index B-V**

## 4.1 Short term variability

From the observations taken at Loiano site of Bologna Astronomical Observatory, no short term variability can be highlighted because the total error is greater than the deviation shown by measurements.

## 4.2 Long term variability

We used Loiano observations and not transformed data from AAVSO database, in filter V, in order to determine the Long Term variability of FY Lac.
It is already possible to see in the preliminary graphic (Fig. 5) a potential Long Term variability of the star.

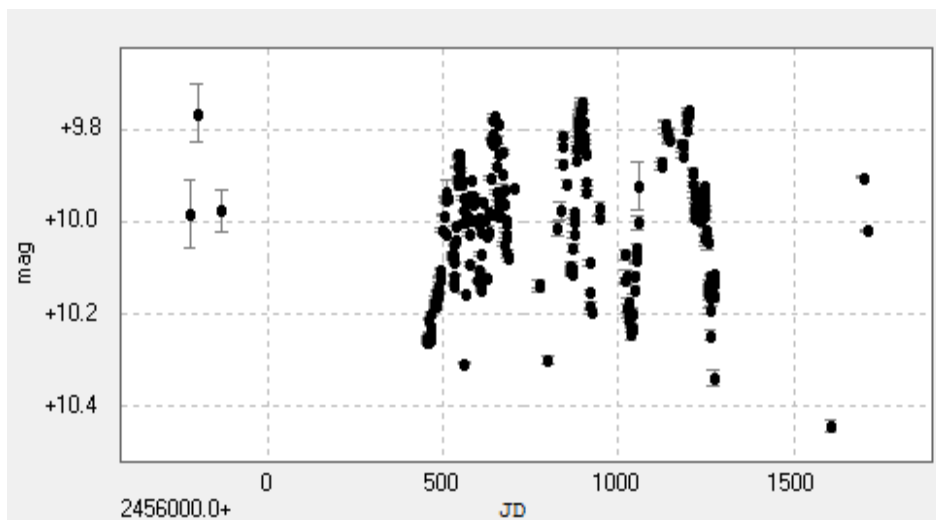

**Fig. 5- Measurements from AAVSO database**

We analyzed the data using Peranso version 2.51, applying Lomb-Scargle and ANOVA period analysis methods.
The following results were found:

- Lomb-Scargle, 271 ± 28 days. Figure 6 shows a phase curve calculated with the AAVSO data and Loiano observations assuming a 271 days period.

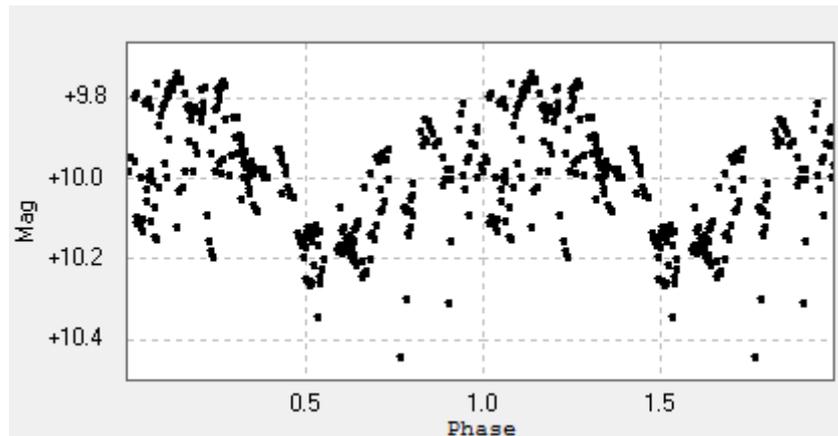

**Fig. 6- Phase curve calculated assuming a 271 days period**

- ANOVA, 277 ± 28 days. Figure 7 shows a phase curve calculated with the AAVSO data and Loiano observations assuming a 277 days period.

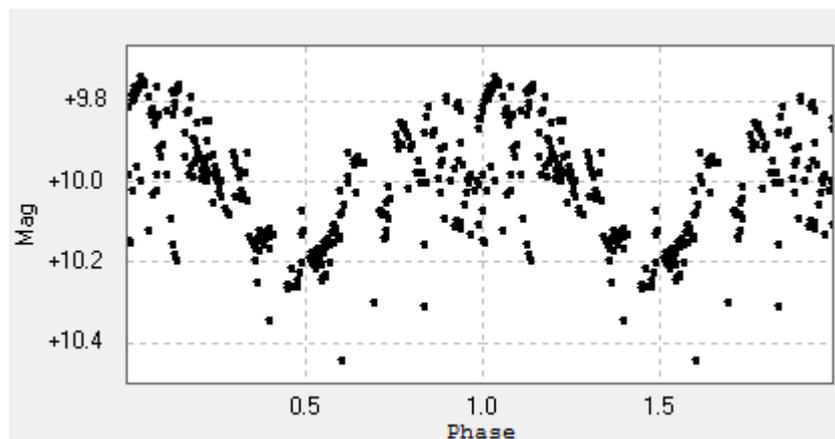

**Fig. 7- Phase curve calculated assuming a 277 days period**

The above analysis shows a preliminary period of 274 ± 28 days.

## 4.3 FY Lac distance and absolute magnitude

The following parallax of FY Lac is available from GAIA DR1 database: 0.496 ± 0.702 mas.
The nominal and minimum distance and absolute magnitude of the star, in V filter, are therefore:
- Nominal distance: 2017 pc
- Minimum distance: 835 pc
- Absolute magnitude (V): -1.43 mag ÷ +0.49 mag.

The value of B-V colour index and the calculated absolute magnitudes are typical of a Red Giant star.
We calculated colour indexes from the mean value of standard magnitude in V filter measured in Loiano and J, H and K magnitudes available from SIMBAD database.
V-J, V-H and V-K colour-indexes were compared (Tab. 4), with stellar fluxes defined in Pickels (1998).

| Spectral type | Log $T_e$ | V-J | V-H | V-K |
|---|---|---|---|---|
|  |  | 4.88 ± 0.28 | 5.85 ± 0.26 | 6.15 ± 0.35 |
| M4 III | 3.551 | 3.95 | 4.88 | 5.07 |
| **M5 III** | **3.534** | **4.66** | **5.62** | **5.84** |
| M6 III | 3.512 | 5.55 | 6.58 | 6.79 |

**Tab. 4- colour indexes vs. spectral types**

FY Lac is therefore compatible with a typical M5 III red giant star fluxes.
We determined a temperature of 3420 °K star using Log $T_e$ data defined in Pickels (1998).

## 5   Conclusions

The analysis of available photometric data showed that FY Lac is a red M5 III pulsating giant star with a preliminary period of 274 ± 28 days.
We also calculated that the temperature of the star is 3420 °K.

Further observations are necessary in order to confirm and refine the preliminary period found for the variable in this report.
Additional observations in R and I filters are also required together with spectroscopic data.
Distance and absolute magnitude will be revised when updated astrometric data will be available.

**Acknowledgements**: This work has made use of data from the European Space Agency (ESA) mission Gaia (https://www.cosmos.esa.int/gaia), processed by the Gaia Data Processing and Analysis Consortium (DPAC, https://www.cosmos.esa.int/web/gaia/dpac/consortium).
We acknowledge with thanks the variable star observations from the AAVSO International Database contributed by observers worldwide and used in this research.
We acknowledge the use of the 1.52m Cassini Telescope run by INAF-Osservatorio Astronomico di Bologna at Loiano site.
This activity has made use of the SIMBAD database, operated at CDS, Strasbourg, France.
This work was carried out in the context of educational and training activities provided by Italian law 'Alternanza Scuola Lavoro', July 13th, 2015 n.107, Art.1, paragraphs 33-43.